\documentclass[sigconf,natbib=true]{acmart}

\usepackage{graphicx}
\usepackage{xspace}
\usepackage{booktabs}
\usepackage{tikz}
\usepackage{enumitem}

\usetikzlibrary{positioning, arrows.meta, calc}

\newcommand{\ours}{\mbox{MMRerank}\xspace}
\providecommand{\Description}[1]{}

\setcopyright{none}
\renewcommand\footnotetextcopyrightpermission[1]{}
\begin{document}

\title{Beyond Relevance: An Empirical Study of Multimodal LLMs for Diversity-Aware Recommendation Reranking}
                                                 
\author{Chengkai Liu}
\affiliation{%
  \institution{Amazon}
  \country{Seattle, WA, USA}}
\email{lchengka@amazon.com}

\author{Yan Fu}
\affiliation{%
  \institution{Amazon}
  \country{Seattle, WA, USA}}
\email{yanfu@amazon.com}

\author{Huiyuan Chen}
\affiliation{%
  \institution{Amazon}
  \country{Sunnyvale, CA, USA}}
\email{huiyuach@amazon.com}

\author{Sheikh Sarwar}
\affiliation{%
  \institution{Amazon}
  \country{Sunnyvale, CA, USA}}
\email{smsarwar@amazon.com}

\author{Qilin Qi}
\affiliation{%
  \institution{Amazon}
  \country{Sunnyvale, CA, USA}}
\email{qilinqi@amazon.com}

\begin{abstract}
Large language models (LLMs) have emerged as promising rerankers for recommendation, yet most existing approaches rely on textual item descriptions and primarily focus on relevance. In visually oriented domains such as video streaming, item images provide complementary signals that may also inform recommendation diversity. This raises two questions: when does visual information improve recommendation quality, and can multimodal LLMs (MLLMs) enhance relevance and diversity simultaneously?
We investigate these questions through a training-free, diversity-aware multimodal reranking framework that jointly incorporates item images, textual metadata, and user interaction histories. We conduct an empirical study across three public Amazon Review datasets and an industrial Amazon Prime Video dataset, using both open-source and proprietary MLLMs. Our results reveal substantial domain-dependent effects: adding visual information yields up to 121.9\% relative improvement in NDCG@5 on Prime Video, while producing limited or inconsistent gains on product recommendation datasets. Furthermore, MLLM-based reranking improves both relevance and category-diversity-sensitive ranking quality over a sequential recommendation baseline, achieving up to 71.0\% improvement in NDCG@5 and 14.1\% in $\alpha$-NDCG@5 on Prime Video.
These findings highlight the conditional value of visual information and the potential of training-free MLLMs for relevance- and diversity-aware recommendation reranking.
\end{abstract}



\maketitle

\section{Introduction}
\label{sec:intro}

Personalized recommendation systems play a central role in content discovery across e-commerce, video streaming, and social media platforms. Beyond identifying relevant items, these systems must also provide diverse recommendations to broaden content exposure and avoid repetitive user experiences~\cite{carbonell1998mmr,wang2020personalized,steck2018calibrated}. This challenge is particularly important in visually oriented domains, where users interact with items through both textual metadata and visual presentations~\cite{he2016vbpr,wei2019mmgcn,liu2024multimodal}. For example, movie cover art and product images convey visual characteristics that may not be captured by titles or categorical labels alone. Importantly, \emph{semantic diversity does not necessarily imply visual diversity}: items from different categories may share similar visual appearances, while items within the same category may exhibit distinct visual styles. This distinction motivates exploring visual information as an additional signal for relevance- and diversity-aware recommendation.

Diversity-aware reranking has been extensively studied through methods based on predefined categories, item representations, and explicit relevance--diversity trade-offs~\cite{chen2018fast,pei2019personalized,chen2022denoising,liu2026rethinking}. While effective, these approaches typically depend on predefined representations or similarity measures, which may not fully capture the visual and semantic characteristics of recommended items. More recently, large language models (LLMs) have demonstrated promising capabilities in recommendation and listwise reranking~\cite{wu2024survey,geng2022recommendation,bao2023tallrec,sun2023chatgpt,hou2024large}. By incorporating user interaction histories and candidate metadata into prompts, LLMs can refine rankings without task-specific training. However, many existing LLM-based rerankers primarily rely on textual descriptions, leaving the potential contribution of visual content to relevance and diversity insufficiently explored.

Multimodal large language models (MLLMs), which jointly process images and text, offer a promising opportunity to address this limitation~\cite{liu2023visual,chen2024internvl}. In principle, visual information may help distinguish items with similar textual descriptions but different appearances, while also revealing similarities that are not reflected in category labels. Nevertheless, incorporating additional modalities does not necessarily guarantee better recommendations: visual signals may be redundant when textual metadata is sufficiently informative, and their utility may vary across domains and model backbones. Moreover, it remains unclear whether MLLM-based reranking can improve recommendation relevance and diversity simultaneously. These considerations motivate three research questions: \textbf{(RQ1)} When does visual information improve recommendation relevance? \textbf{(RQ2)} Can MLLM-based reranking simultaneously improve relevance and category-diversity-sensitive ranking quality? \textbf{(RQ3)} How do these effects vary across recommendation domains and MLLM backbones?

To investigate these questions, we study \ours, a training-free multimodal reranking framework that integrates item images, textual metadata, and chronological user interaction histories through structured prompts. Specifically, item images are organized into composite visual inputs and aligned with their corresponding titles, allowing pretrained MLLMs to consider visual and semantic information jointly. The prompts further incorporate explicit instructions to balance user relevance with recommendation diversity. We examine the framework in two settings: standalone recommendation over sampled candidates and reranking of candidate lists produced by a sequential recommender such as SASRec~\cite{kang2018self}. This setup enables us to investigate the contribution of visual information and the behavior of MLLM-based reranking without task-specific model training.

We conduct experiments across three public Amazon Review datasets and an industrial Amazon Prime Video dataset, using both open-source and proprietary MLLMs. Our findings reveal substantial domain-dependent differences in the value of visual information. On Prime Video, incorporating images improves NDCG@5 by up to 121.9\% relative to text-only prompting, whereas improvements on product recommendation datasets are limited or inconsistent. Furthermore, MLLM-based reranking can simultaneously improve relevance and category-diversity-sensitive ranking quality over a sequential recommendation baseline, achieving up to 71.0\% higher NDCG@5 and 14.1\% higher $\alpha$-NDCG@5 on Prime Video. These results highlight both the potential and the limitations of multimodal LLMs for diversity-aware recommendation. 

Our contributions are summarized as follows:
\begin{itemize}[leftmargin=1.2em, labelsep=0.4em, nosep]
    \item We investigate training-free multimodal LLM reranking from the joint perspective of relevance and diversity, using a lightweight framework that integrates visual and textual information with diversity-aware prompting.
    \item We provide empirical evidence that MLLM-based reranking can simultaneously improve relevance and category-diversity-sensitive ranking quality over a sequential recommender.
    \item We identify substantial domain- and backbone-dependent effects of visual information, offering insights into when multimodal reranking is beneficial.
\end{itemize}

\section{Related Work}
\label{sec:related}

\noindent\textbf{Multimodal Recommendation.}
Multimodal recommendation has been widely studied to enrich item representations and improve recommendation relevance~\cite{liu2024multimodal}. Early approaches incorporate visual features into collaborative filtering or graph-based models, including VBPR~\cite{he2016vbpr}, MMGCN~\cite{wei2019mmgcn}, and BM3~\cite{zhou2023bootstrap}. More recently, multimodal large language models (MLLMs) have enabled joint reasoning over visual and textual information~\cite{alayrac2022flamingo,li2023blip,liu2023visual}. Methods such as UniMP~\cite{wei2024towards} and MLLM-MSR~\cite{ye2025mllmmsr} further adapt MLLMs to recommendation through task-specific training. In contrast, we investigate pretrained MLLMs as training-free rerankers, focusing on when visual information provides additional value for relevance and diversity-sensitive ranking quality.

\noindent\textbf{LLM-based Reranking.}
Large language models have demonstrated promising capabilities in recommendation~\cite{wu2024survey,geng2022recommendation,bao2023tallrec}, including zero-shot listwise reranking~\cite{sun2023chatgpt,hou2024large}. Recent approaches further incorporate diversity-aware instructions into LLM-based reranking~\cite{carraro2024enhancing,gao2025llm4rerank}, while benchmarks such as MSRBench~\cite{he2025msrbench} investigate vision-language models for sequential recommendation. However, the role of visual inputs in diversity-aware reranking remains insufficiently characterized. Rather than introducing a new reranking algorithm, our work empirically examines whether multimodal prompting improves relevance and diversity-sensitive ranking quality, and how these effects vary across domains and MLLM backbones.

\noindent\textbf{Diversity-Aware Recommendation.}
Diversity-aware recommendation seeks to balance relevance with exposure to varied content. Classical approaches include maximal marginal relevance~\cite{carbonell1998mmr}, determinantal point processes~\cite{chen2018fast}, and calibration-based methods~\cite{steck2018calibrated,zhao2024can,chen2021structured}, which typically rely on predefined categories, item similarities, or explicit relevance--diversity trade-offs~\cite{pei2019personalized,wang2020personalized}. However, category-level diversity may not fully reflect the visual variety of recommended items, particularly in image-centric interfaces. This motivates investigating MLLMs that jointly consider item appearance and semantic information. Our study evaluates their ability to improve relevance and category-diversity-sensitive ranking quality, while recognizing visual diversity as a complementary dimension beyond the scope of our current evaluation.

\begin{figure*}[t]
    \centering
    \includegraphics[width=\textwidth]{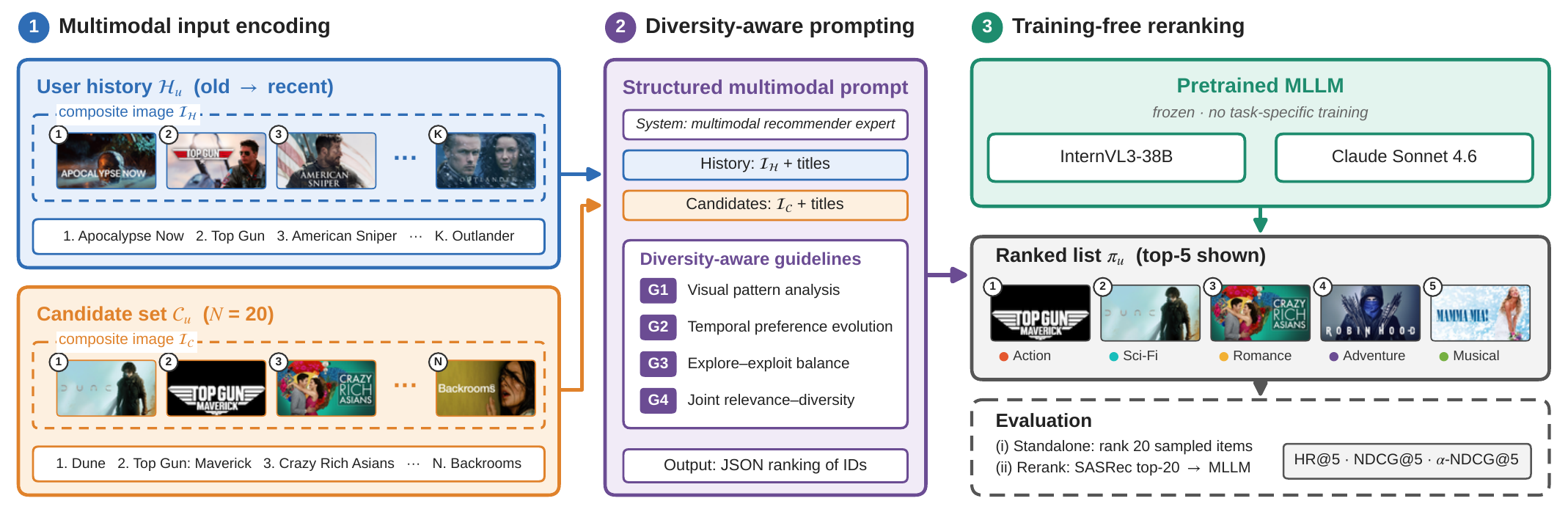}
    \caption{Overview of \ours. Chronological user histories and candidate items are encoded as composite images aligned with numbered titles. Diversity-aware instructions are combined with these inputs in a structured prompt, and a pretrained MLLM produces a ranked permutation of the candidates without task-specific training.}
    \Description{Three-stage MMRerank framework showing multimodal input encoding of user histories and candidate items, diversity-aware prompting, and training-free MLLM ranking.}
    \label{fig:framework}
\end{figure*}

\section{Methodology}
\label{sec:method}

We investigate multimodal LLMs for relevance- and diversity-aware recommendation through \ours, a training-free reranking framework. As illustrated in Figure~\ref{fig:framework}, \ours integrates visual and textual item information with user interaction histories through structured multimodal prompting. Rather than introducing a task-specific ranking model or an explicit diversity optimization algorithm, the framework provides a unified setting for examining how pretrained MLLMs utilize visual information and respond to relevance--diversity ranking instructions.

\subsection{Problem Formulation}

Let $u$ denote a user with a chronological interaction history $\mathcal{H}_u = \{(t_i, v_i)\}_{i=1}^{K}$, where $t_i$ and $v_i$ represent the title and image of the $i$-th interacted item, respectively. Interactions are ordered from oldest to most recent, allowing the model to observe both long-term preferences and recent interests. Given a candidate set $\mathcal{C}_u = \{c_1, \ldots, c_N\}$, where each candidate $c_j$ is associated with a title $t(c_j)$ and an image $v(c_j)$, the objective is to produce a ranked permutation of these candidates that considers both relevance to the user's preferences and diversity across recommended items.

Conventional diversity-aware reranking often formulates this objective as a weighted combination of relevance and diversity scores. For example, maximal marginal relevance (MMR)~\cite{carbonell1998mmr} greedily selects
\begin{equation}
    c^{*}=\arg\max_{c\in\mathcal{C}_u\setminus S}\ \lambda\,\mathrm{rel}(c\mid\mathcal{H}_u)-(1-\lambda)\max_{c'\in S}\mathrm{sim}(c,c'),
    \label{eq:mmr}
\end{equation}
where $S$ is the set of already selected items, $\mathrm{rel}(\cdot)$ is a relevance score, $\mathrm{sim}(\cdot,\cdot)$ is a predefined item similarity, and $\lambda$ is an explicit trade-off parameter~\cite{wang2020personalized}. In contrast, we investigate whether a pretrained MLLM can perform relevance- and diversity-aware reranking directly from multimodal context and natural-language instructions. Specifically, we define the generated ranking as
\begin{equation}
    \pi_u =
    f_{\mathrm{MLLM}}
    \left(
        \mathcal{H}_u,
        \mathcal{C}_u;
        \mathcal{P}
    \right),
    \label{eq:rerank}
\end{equation}
where $\pi_u$ is a permutation of candidate indices, $f_{\mathrm{MLLM}}$ denotes the pretrained multimodal language model, and $\mathcal{P}=(p_{\mathrm{sys}},\mathcal{G},p_{\mathrm{out}})$ specifies the structured prompt, consisting of a system instruction, the ranking guidelines $\mathcal{G}=\{g_1,\ldots,g_4\}$ (Section~\ref{sec:diversity}), and an output specification. Eq.~\eqref{eq:rerank} is instantiated in Eqs.~\eqref{eq:input} and~\eqref{eq:fallback}.
Compared with Eq.~\eqref{eq:mmr}, $\mathrm{rel}(\cdot)$, $\mathrm{sim}(\cdot,\cdot)$, and $\lambda$ are not specified explicitly; they are only implicitly conveyed through $\mathcal{P}$ and the multimodal inputs.
This formulation does not assume that the MLLM explicitly optimizes a known relevance--diversity objective. Instead, the model generates a candidate ordering conditioned on user history, item content, and diversity-aware instructions. The resulting ranking is subsequently evaluated using relevance and category-diversity-sensitive metrics. Importantly, the same framework supports both standalone recommendation over sampled candidates and reranking of candidate lists produced by a sequential recommender, enabling controlled comparisons of multimodal and text-only prompting.

\subsection{Multimodal Prompt Construction}

\begin{figure}[t]
\centering
\fbox{\begin{minipage}{0.92\columnwidth}\small
\textbf{[System]} You are a multimodal recommender system expert.\\[3pt]
\textbf{[User]}\\
\textit{History:} \texttt{[Image:} $\mathcal{I}_\mathcal{H}$\texttt{]} {\scriptsize (concatenated, chronological L$\to$R)}\\
\quad \texttt{1.~\{title$_1$\}~~2.~\{title$_2$\}~~$\cdots$~~K.~\{title$_K$\}}\\[3pt]
\textit{Candidates:} \texttt{[Image:} $\mathcal{I}_\mathcal{C}$\texttt{]}\\
\quad \texttt{1.~\{cand$_1$\}~~2.~\{cand$_2$\}~~$\cdots$~~N.~\{cand$_N$\}}\\[3pt]
\textit{Guidelines:}\\
\quad 1. Analyze \textbf{visual patterns} (style, color, design) and metadata\\
\quad 2. Consider chronological evolution of user preferences\\
\quad 3. Balance similarity with potential new interests\\
\quad 4. Maximize both \textbf{relevance} and \textbf{genre diversity}\\[2pt]
\textit{Output:} \texttt{\{``recommendations'': [id$_1$, $\ldots$, id$_N$]\}}
\end{minipage}}
\caption{Simplified prompt template used in \ours (reranker setting). The prompt encodes multimodal history and candidates as composite images with numbered titles, and provides structured guidelines for joint relevance--diversity optimization.}
\Description{Structured prompt template showing system instruction, multimodal user history with composite image and numbered titles, candidate encoding, and task instruction for relevance and diversity ranking.}
\label{fig:prompt}
\end{figure}

A central component of \ours is the construction of a structured multimodal prompt that presents user interaction histories and candidate items in a unified visual--textual format. Unlike text-only reranking, which represents items primarily through titles or descriptions, our framework exposes the MLLM to both item appearance and semantic information. As illustrated in Figure~\ref{fig:prompt}, the prompt consists of four components: composite visual encoding, cross-modal alignment, candidate representation, and structured ranking instructions.

\noindent\textbf{Visual encoding via composite images.}
Let $\rho(\cdot)$ resize an image to a common resolution $H\times W$. Given the $K$ most recent interactions and the $N$ candidates, we concatenate the resized images horizontally to construct two composite images,
\begin{equation}
\begin{aligned}
    \mathcal{I}_{\mathcal{H}} &=
    \mathrm{Concat}\big(\rho(v_1),\ldots,\rho(v_K)\big), \\
    \mathcal{I}_{\mathcal{C}} &=
    \mathrm{Concat}\big(\rho(v(c_1)),\ldots,\rho(v(c_N))\big),
\end{aligned}
\label{eq:composite}
\end{equation}
with $\mathcal{I}_{\mathcal{H}}\in\mathbb{R}^{H\times (KW)\times 3}$ and $\mathcal{I}_{\mathcal{C}}\in\mathbb{R}^{H\times (NW)\times 3}$. In $\mathcal{I}_{\mathcal{H}}$, the images are arranged chronologically from left to right, with the most recent interaction appearing last. This representation serves three purposes. First, it consolidates multiple item images into a single visual input, reducing the number of separately supplied images and simplifying prompt construction. Second, it places historical items within a shared visual context, allowing the MLLM to examine recurring or contrasting visual characteristics, such as color palettes, artistic styles, and composition, across the user's interactions. Third, the spatial arrangement preserves temporal order, providing a consistent visual representation of how the user's consumed content evolves over time. Although the composite format reduces the number of image inputs, its actual visual-token cost depends on the MLLM's image preprocessing and resolution handling.

\noindent\textbf{Textual encoding and cross-modal alignment.}
Visual appearance alone may be insufficient to characterize an item's semantic content. For example, visually similar movie posters may correspond to different genres or narratives, while products with similar titles may differ substantially in appearance. To provide complementary semantic information, we pair each composite image with a numbered list of item titles, following the same ordering as the visual segments. Specifically, the $i$-th title corresponds to the $i$-th image in the composite, establishing an explicit alignment between textual and visual representations. This numbering allows the MLLM to associate each item's appearance with its title and to interpret visual patterns in the context of semantic information. For user histories, the shared ordering additionally connects item content with interaction chronology, supporting the interpretation of both recurring preferences and recent interests.

\noindent\textbf{Candidate representation.}
We apply the same encoding strategy to the candidate set $\mathcal{C}_u$: the composite image $\mathcal{I}_{\mathcal{C}}$ in Eq.~\eqref{eq:composite} is accompanied by a numbered list of candidate titles. The candidate indices serve as stable identifiers throughout the reranking process, allowing the model to refer to items without reproducing their full titles. Importantly, history and candidates are represented separately: the former provides evidence of user preferences, while the latter defines the items to be ranked. Using a consistent visual--textual representation for both groups allows the MLLM to compare candidate characteristics against historical interests while considering similarities and differences among candidates.

Formally, let $\mathcal{T}_{\mathcal{H}}=(t_1,\ldots,t_K)$ and $\mathcal{T}_{\mathcal{C}}=(t(c_1),\ldots,t(c_N))$ be the numbered title lists of the history and the candidates. The $i$-th title is aligned with the $i$-th column block $[(i-1)W+1,\,iW]$ of the corresponding composite image, and the full multimodal input and the raw model output are
\begin{equation}
    x_u=\Phi\big(\mathcal{I}_{\mathcal{H}},\mathcal{T}_{\mathcal{H}},\mathcal{I}_{\mathcal{C}},\mathcal{T}_{\mathcal{C}};\mathcal{P}\big),\qquad o_u=f_{\mathrm{MLLM}}(x_u),
    \label{eq:input}
\end{equation}
where $\Phi$ assembles the prompt template in Figure~\ref{fig:prompt}. In the title-only setting, $\mathcal{I}_{\mathcal{H}}$ and $\mathcal{I}_{\mathcal{C}}$ are removed from $\Phi$ and the remaining prompt is unchanged.

\noindent\textbf{Prompt assembly and output parsing.}
The complete prompt follows a structured format comprising (i) a system instruction defining the recommendation task, (ii) the multimodal user history, (iii) the multimodal candidate set, (iv) explicit guidelines for balancing relevance and diversity (Section~\ref{sec:diversity}), and (v) an output specification. Rather than requesting free-form recommendations, we instruct the MLLM to return a JSON object containing an ordered list of candidate indices. This constrains the task to reordering the supplied candidates and facilitates automatic evaluation.
Let $\mathrm{Parse}(\cdot)$ extract the index list from the returned JSON object, and let $\mathrm{Perm}(\mathcal{C}_u)$ denote the set of all permutations of the $N$ candidate indices. An output $o_u$ is valid if $\mathrm{Parse}(o_u)\in\mathrm{Perm}(\mathcal{C}_u)$. If $o_u$ is invalid, the model is prompted once more to obtain $o_u'$; if parsing still fails, we retain the original candidate ordering $\pi_u^{\mathrm{base}}$ (the SASRec order in the reranking setting). The final ranking is therefore
\begin{equation}
    \pi_u=
    \begin{cases}
        \mathrm{Parse}(o_u) & \text{if } \mathrm{Parse}(o_u)\in\mathrm{Perm}(\mathcal{C}_u),\\
        \mathrm{Parse}(o_u') & \text{else if } \mathrm{Parse}(o_u')\in\mathrm{Perm}(\mathcal{C}_u),\\
        \pi_u^{\mathrm{base}} & \text{otherwise.}
    \end{cases}
    \label{eq:fallback}
\end{equation}
The same prompt structure is applied across open-source and proprietary MLLMs, enabling comparisons under a consistent input and instruction format.

\subsection{Diversity-Aware Ranking Instructions}
\label{sec:diversity}

Beyond incorporating multimodal inputs, \ours explicitly introduces relevance and diversity considerations into the ranking task through natural-language instructions. Unlike conventional methods that combine relevance and similarity scores with an explicit trade-off parameter (Eq.~\eqref{eq:mmr}), our framework requires neither an explicit diversity score nor a manually specified coefficient. Instead, we investigate whether pretrained MLLMs can incorporate multiple ranking considerations directly through structured prompting. Specifically, we design four complementary guidelines $\mathcal{G}=\{g_1,\ldots,g_4\}$ that encourage the model to examine visual characteristics, interpret user preferences, consider exploratory recommendations, and balance relevance with diversity.

\noindent\textbf{(G1) Visual pattern analysis.}
The first guideline $g_1$ instructs the MLLM to examine visual characteristics of both historical and candidate items, including color palettes, artistic styles, composition, and other appearance-related attributes, alongside textual metadata. This instruction is motivated by the observation that semantic descriptions and visual appearances capture different aspects of item content. For instance, two movies categorized as action thrillers may exhibit substantially different visual styles, while movies from different genres may share similar poster designs. By explicitly directing attention to visual information, we aim to encourage the model to use item images as complementary evidence rather than relying exclusively on titles or genre-related semantic cues. This guideline also provides a basis for comparing multimodal and text-only prompting, allowing us to examine whether access to visual information improves ranking quality.

\noindent\textbf{(G2) Temporal preference evolution.}
User preferences may evolve over time, and recent interactions can provide information that differs from long-term consumption patterns. Accordingly, the second guideline $g_2$ asks the MLLM to consider the chronological progression of the user's interaction history when interpreting preferences. As described in Section~\ref{sec:method}, historical images are arranged from oldest to most recent, with corresponding titles following the same ordering. This representation allows the prompt to communicate temporal context without introducing a separate temporal modeling component. For example, a user may have historically consumed action movies but recently shown greater interest in comedy or family-oriented content. The guideline encourages the MLLM to consider such changes when evaluating candidates, rather than treating the interaction history as an unordered collection of preferences.

\noindent\textbf{(G3) Exploration--exploitation balance.}
A recommendation list consisting entirely of items similar to a user's past interactions may provide high immediate relevance but limited variety. Conversely, excessive exploration may introduce items that are insufficiently aligned with the user's interests. The third guideline $g_3$ therefore encourages the MLLM to balance familiar content with potentially relevant items that extend beyond previously observed preferences. Rather than treating exploration as random diversification, the instruction emphasizes identifying candidates that may introduce new interests while remaining consistent with the user's broader consumption history. This provides a qualitative mechanism for encouraging diversity without requiring an explicit exploration probability or a predefined similarity threshold.

\noindent\textbf{(G4) Joint relevance--diversity consideration.}
The final guideline $g_4$ explicitly instructs the MLLM to consider both recommendation relevance and genre or category diversity when producing the ranked list. Relevance is assessed in the context of the user's interaction history, while diversity encourages coverage of different content categories among highly ranked candidates. Unlike a post-hoc diversification procedure that first constructs a relevance-based ranking and subsequently adjusts it using a separate diversity score, our framework presents these considerations within a single ranking instruction. The MLLM is therefore asked to generate one candidate ordering that accounts for both objectives, without an explicit mathematical optimization step or a fixed relevance--diversity weighting parameter.

Together, these guidelines define a consistent diversity-aware prompting protocol across the evaluated MLLMs and recommendation domains. Their purpose is to investigate whether pretrained models can use multimodal context and natural-language instructions to produce rankings that perform well under both relevance and diversity-sensitive evaluation. Importantly, the instructions do not guarantee that the model optimizes either objective or learns an optimal user-specific relevance--diversity trade-off. We therefore assess the resulting rankings empirically, using relevance metrics and category-based diversity-sensitive metrics, rather than assuming that the intended prompting behavior is achieved.

\section{Experiments}
\label{sec:exp}

\subsection{Experimental Setup}
\label{sec:exp-setup}
\noindent\textbf{Datasets.}
We conduct experiments on four datasets spanning two recommendation domains.
(1)~\textbf{Amazon Review}~\cite{he2016vbpr}: three public product recommendation datasets (\textit{Beauty}, \textit{Sports}, and \textit{Toys}), containing user interactions, item titles, and product images.
(2)~\textbf{Amazon Prime Video}: a proprietary video streaming dataset containing user interaction histories, content titles, and cover art, providing a complementary evaluation setting for visually oriented recommendation.
For the public Amazon Review datasets, we retain users with at least five interactions and evaluate 200 sampled users per dataset.
For Prime Video, we report only relative performance changes to avoid disclosing proprietary evaluation statistics.
We use up to the $K=10$ most recent interactions as user history.

\noindent\textbf{Evaluation Protocol.}
We evaluate two settings with $N=20$ candidates.
(i)~\emph{Standalone recommendation}: the candidate set comprises the ground-truth item and 19 randomly sampled negatives, and the MLLM ranks these candidates directly.
(ii)~\emph{Reranking}: the candidate set consists of the top-20 items retrieved by SASRec~\cite{kang2018self} from the full item catalog, and the MLLM reorders this list. Because the ground-truth item may not be retrieved, reranking can only improve ranking quality within SASRec's top-20, and SASRec's HR@20 bounds the achievable HR@5.
Both settings use the same multimodal prompt construction and diversity-aware ranking instructions described in Section~\ref{sec:method}.
Results are therefore not directly comparable across the two settings: the standalone setting measures ranking ability over sampled candidates, whereas the reranking setting measures the incremental value of MLLM reranking on top of a sequential recommender.

\noindent\textbf{Models and Baselines.}
We adopt SASRec~\cite{kang2018self}, a self-attentive sequential recommendation model~\cite{hidasi2016session,sun2019bert4rec}, as the base recommender for reranking.
We evaluate two pretrained MLLMs: \textbf{InternVL3-38B}~\cite{zhu2025internvl3}, an open-source model, and \textbf{Claude Sonnet 4.6}~\cite{anthropic2024claude}, a proprietary API-based model.
\footnote{Claude Sonnet 4.6 was the latest available model at the time of experimentation. Our experimental framework is model-agnostic and can be applied to newer MLLMs without modifying the reranking procedure. However, the generalizability of our empirical findings to newer models remains to be evaluated.}
Both models follow the same prompt structure.
We compare \emph{title-only} and \emph{title+image} inputs to examine the contribution of visual information, while SASRec provides the baseline for evaluating the incremental value of MLLM reranking.

\noindent\textbf{Evaluation Metrics.}
We report Hit Ratio (HR@5) and Normalized Discounted Cumulative Gain (NDCG@5) to measure recommendation relevance.
To evaluate category diversity, we additionally report $\alpha$-NDCG@5~\cite{clarke2008novelty} with $\alpha=0.5$, using item genres or categories as aspects. Each candidate is considered relevant to the categories it belongs to, independent of the ground-truth item, so $\alpha$-NDCG@5 measures position-discounted category coverage among the top-ranked items, normalized by the ideal ordering of the same candidate set. NDCG@5 and $\alpha$-NDCG@5 therefore capture complementary aspects of ranking quality: relevance to the held-out item and category diversity, respectively.

\subsection{RQ1: When Does Visual Information Improve Recommendation?}
\label{sec:visual-impact}

We first investigate the contribution of visual information by comparing title-only and title+image prompting in the standalone recommendation setting. Tables~\ref{tab:amazon-results} and~\ref{tab:pv} report the results on the three Amazon Review datasets and the Prime Video dataset, respectively. Across these domains, we observe substantial differences in the benefits of incorporating item images.

\noindent\textbf{Limited and inconsistent gains on product recommendation.}
As shown in Table~\ref{tab:amazon-results}, adding images produces relatively small and mixed changes on the Amazon Review datasets. For InternVL3-38B, visual inputs decrease NDCG@5 on Beauty (0.4727$\to$0.4535) and Sports (0.4861$\to$0.4588), while leaving performance nearly unchanged on Toys (0.5091$\to$0.5093). Claude Sonnet 4.6 exhibits a similar pattern of limited visual gains, with modest NDCG@5 improvements on Sports and Toys but a slight decrease on Beauty. These results suggest that visual information does not consistently provide additional ranking value in the evaluated product recommendation settings. One possible explanation is that product titles already contain informative semantic attributes, leaving less complementary information for images to contribute.

\noindent\textbf{Substantial improvements on video recommendation.}
In contrast, visual inputs yield considerably larger gains on Prime Video (Table~\ref{tab:pv}). For InternVL3-38B, adding cover images improves HR@5 by 65.5\% and NDCG@5 by 121.9\% relative to title-only prompting. Claude Sonnet 4.6 also benefits from visual information, achieving improvements of 14.7\% in HR@5 and 9.6\% in NDCG@5. The improvements are particularly pronounced for InternVL3-38B, indicating that the magnitude of visual gains depends not only on the recommendation domain but also on the MLLM backbone. A plausible explanation is that movie and television titles alone may provide limited information about visual style, tone, or content characteristics, whereas cover art offers complementary cues for matching candidates to user preferences.

\noindent\textbf{Effects on diversity-sensitive ranking quality.}
We additionally examine whether visual inputs affect $\alpha$-NDCG@5, which measures category coverage among the top-ranked items. On Prime Video, adding images improves $\alpha$-NDCG@5 by 15.4\% for InternVL3-38B and 2.4\% for Claude Sonnet 4.6. However, the corresponding changes on Amazon Review are smaller and not uniformly positive. These observations indicate that visual information can improve category-diversity-sensitive ranking quality in some settings, although the metric does not directly establish improvements in visual diversity.

Overall, our findings reveal a pronounced domain-dependent effect of visual information. While multimodal prompting offers limited or inconsistent benefits on the evaluated product datasets, it substantially improves recommendation relevance on Prime Video. Thus, incorporating images into MLLM-based recommendation is not universally beneficial; its effectiveness depends on the interaction between item content, available textual information, and model capabilities.

\begin{table}[t]
\centering
\caption{MLLM as recommender on Amazon datasets. Top-5 metrics for title-only vs.\ title+image inputs. Best result per dataset per metric is \textbf{bolded}.}
\label{tab:amazon-results}
\resizebox{\columnwidth}{!}{%
\begin{tabular}{lccc|ccc}
\toprule
\textbf{Metric (Top@5)} & \multicolumn{3}{c|}{\textbf{InternVL3-38B}} & \multicolumn{3}{c}{\textbf{Claude Sonnet 4.6}} \\
 & Beauty & Sports & Toys & Beauty & Sports & Toys \\
\midrule
HR@5 (title) & 0.6250 & 0.6250 & 0.6050 & 0.6650 & \textbf{0.7200} & \textbf{0.6750} \\
NDCG@5 (title) & 0.4727 & 0.4861 & 0.5091 & \textbf{0.5193} & 0.5100 & 0.5304 \\
$\alpha$-NDCG@5 (title) & 0.8946 & 0.9119 & 0.8773 & 0.8990 & 0.9093 & 0.8584 \\
\midrule
HR@5 (title+img) & 0.6100 & 0.6150 & 0.6250 & \textbf{0.6900} & \textbf{0.7200} & \textbf{0.6750} \\
NDCG@5 (title+img) & 0.4535 & 0.4588 & 0.5093 & 0.5165 & \textbf{0.5173} & \textbf{0.5402} \\
$\alpha$-NDCG@5 (title+img) & 0.8825 & \textbf{0.9186} & \textbf{0.8851} & \textbf{0.9032} & 0.9132 & 0.8802 \\
\bottomrule
\end{tabular}}
\end{table}

\begin{table}[t]
\centering\small
\caption{Relative improvement (\%) on the proprietary Prime Video dataset. \emph{Visual}: title+image vs.\ title-only (standalone). \emph{Rerank}: \ours vs.\ SASRec.}
\label{tab:pv}
\begin{tabular}{lcccc}
\toprule
 & \multicolumn{2}{c}{\textbf{Visual input}} & \multicolumn{2}{c}{\textbf{Reranking}} \\
\cmidrule(lr){2-3}\cmidrule(lr){4-5}
\textbf{Metric} & InternVL3 & Sonnet 4.6 & InternVL3 & Sonnet 4.6 \\
\midrule
HR@5            & \textbf{+65.5}  & +14.7 & \textbf{+52.3} & +48.2 \\
NDCG@5          & \textbf{+121.9} & +9.6  & +66.4 & \textbf{+71.0} \\
$\alpha$-NDCG@5 & \textbf{+15.4}  & +2.4  & +12.0 & \textbf{+14.1} \\
\bottomrule
\end{tabular}
\end{table}

\subsection{RQ2: Can MLLMs Improve Relevance and Diversity Simultaneously?}
\label{sec:relevance-diversity}

We next examine whether MLLM-based reranking can improve recommendation relevance and category diversity simultaneously.
Unlike RQ1, which isolates the contribution of visual inputs over sampled candidates, RQ2 evaluates the complete multimodal
reranking framework on candidate lists retrieved by SASRec.
Table~\ref{tab:amazon-results-rerank} reports absolute ranking metrics on Amazon Review, while Table~\ref{tab:pv} reports relative improvements on Prime Video.

\noindent\textbf{Reranking effectiveness on product recommendation.}
Table~\ref{tab:amazon-results-rerank} shows that MLLM reranking can substantially improve ranking relevance over SASRec, although the magnitude and consistency of the gains vary across models and product categories. Claude Sonnet 4.6 improves both HR@5 and NDCG@5 on all three Amazon Review datasets, achieving its largest relative gains on Sports: HR@5 increases from 0.2500 to 0.6000 (+140.0\%), while NDCG@5 increases from 0.1880 to 0.4285 (+127.9\%). InternVL3-38B also improves relevance on Sports and Toys, but performs worse than SASRec on Beauty, where NDCG@5 decreases from 0.3461 to 0.3033. These results demonstrate that the benefits of MLLM reranking are not uniform across backbones and recommendation categories.

Importantly, relevance improvements can be accompanied by higher diversity-sensitive ranking quality. For Claude Sonnet 4.6, $\alpha$-NDCG@5 improves across all three product datasets, including Beauty (0.9135$\to$0.9481), Sports (0.9214$\to$0.9496), and Toys (0.8785$\to$0.9319). InternVL3-38B exhibits a less consistent pattern, with improvements on Toys but decreases on Beauty and Sports. Thus, while MLLM reranking can improve relevance and category-diversity-sensitive quality together, this behavior is not guaranteed for every model or dataset.

\noindent\textbf{Consistent gains on Prime Video.}
The results on Prime Video (Table~\ref{tab:pv}) show a more consistent pattern. Both MLLMs improve all three metrics over the SASRec baseline. InternVL3-38B achieves relative gains of 52.3\% in HR@5, 66.4\% in NDCG@5, and 12.0\% in $\alpha$-NDCG@5. Claude Sonnet 4.6 achieves corresponding improvements of 48.2\%, 71.0\%, and 14.1\%. These results provide evidence that multimodal MLLM reranking can improve relevance and category-diversity-sensitive ranking quality simultaneously, particularly in the evaluated video recommendation setting.

\begin{figure}[t]
    \centering
    \includegraphics[width=\columnwidth]{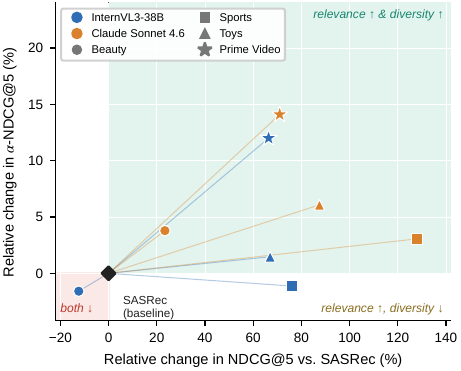}
    \caption{Relative changes in NDCG@5 and $\alpha$-NDCG@5 after MLLM reranking compared with SASRec across four datasets and two model backbones. Each point represents a dataset--model pair. The upper-right quadrant indicates simultaneous improvements in relevance and diversity-sensitive ranking quality.}
    \Description{Scatter plot of relative NDCG and alpha-NDCG changes for InternVL3-38B and Claude Sonnet 4.6 across Beauty, Sports, Toys, and Prime Video.}
    \label{fig:relevance-diversity}
\end{figure}

\noindent\textbf{Interpreting relevance--diversity improvements.}
Figure~\ref{fig:relevance-diversity} visualizes the joint changes in NDCG@5 and $\alpha$-NDCG@5 across datasets and MLLM backbones, highlighting both simultaneous improvements and mixed outcomes. Because $\alpha$-NDCG@5 is computed from category labels independently of the ground-truth item, gains in both metrics indicate that relevance improvements are not obtained by concentrating top-ranked items in fewer categories. These results should nevertheless be interpreted with care. Category coverage does not capture visual diversity, and $\alpha$-NDCG@5 depends on the category composition of the candidate set. Moreover, our comparisons evaluate the complete multimodal reranking framework rather than isolating the effect of diversity-aware instructions.

Taken together, these findings suggest that pretrained MLLMs offer a promising training-free approach to relevance- and diversity-aware reranking. Their effectiveness is strongest and most consistent on Prime Video, while results on the product datasets reveal meaningful differences across model backbones and categories.

\begin{table}[t]
\centering
\caption{\ours as reranker on Amazon datasets. Top-5 metrics before (SASRec baseline) and after MLLM reranking. Best reranked result per dataset is \textbf{bolded}.}
\label{tab:amazon-results-rerank}
\resizebox{\columnwidth}{!}{%
\begin{tabular}{lccc|ccc}
\toprule
\textbf{Metric (Top@5)} & \multicolumn{3}{c|}{\textbf{InternVL3-38B}} & \multicolumn{3}{c}{\textbf{Claude Sonnet 4.6}} \\
 & Beauty & Sports & Toys & Beauty & Sports & Toys \\
\midrule
HR@5 & 0.4650 & 0.2500 & 0.3550 & 0.4650 & 0.2500 & 0.3550 \\
NDCG@5 & 0.3461 & 0.1880 & 0.2370 & 0.3461 & 0.1880 & 0.2370 \\
$\alpha$-NDCG@5 & 0.9135 & 0.9214 & 0.8785 & 0.9135 & 0.9214 & 0.8785 \\
\midrule
HR@5 (Rerank) & 0.3950 & 0.4550 & 0.4750 & \textbf{0.5650} & \textbf{0.6000} & \textbf{0.5450} \\
NDCG@5 (Rerank) & 0.3033 & 0.3309 & 0.3959 & \textbf{0.4270} & \textbf{0.4285} & \textbf{0.4444} \\
$\alpha$-NDCG@5 (Rerank) & 0.8990 & 0.9110 & 0.8914 & \textbf{0.9481} & \textbf{0.9496} & \textbf{0.9319} \\
\bottomrule
\end{tabular}}
\end{table}

\subsection{RQ3: How Do Effects Vary Across Domains and Backbones?}
\label{sec:cross-domain}

Our results reveal that the effectiveness of multimodal recommendation depends jointly on the recommendation domain and the MLLM backbone. Beyond the individual findings of RQ1 and RQ2, we highlight two broader observations.

\noindent\textbf{Visual gains do not necessarily imply stronger ranking performance.}
On Prime Video, InternVL3-38B achieves substantially larger relative NDCG@5 gains from visual inputs than Claude Sonnet 4.6 (121.9\% vs.\ 9.6\%). However, Claude achieves a slightly larger improvement over SASRec in the reranking setting (71.0\% vs.\ 66.4\%). These results illustrate that the marginal benefit of visual information and the overall effectiveness of MLLM reranking are distinct properties. A larger relative improvement from images does not necessarily translate into a stronger final reranker.

\noindent\textbf{Multimodal benefits are conditional rather than universal.}
Neither MLLM consistently benefits from visual inputs across all three Amazon Review datasets, despite the substantial improvements observed on Prime Video. Moreover, the consistency of reranking gains differs across backbones: Claude improves relevance and $\alpha$-NDCG@5 across all three product datasets, whereas InternVL3-38B exhibits mixed results. These observations suggest that multimodal reranking should be evaluated jointly across domains, input modalities, and model backbones, rather than assuming that additional visual information or stronger general-purpose models will always improve recommendation quality.

\subsection{Inference Cost and Practical Considerations}
\label{sec:efficiency}

We examine the inference cost of \ours using InternVL3-38B with 4-bit quantization. The model runs on a single NVIDIA A100 40GB GPU, requiring less than 30GB of GPU memory and approximately 30 seconds to process each user. These measurements highlight the computational overhead of applying large MLLMs to recommendation, even when reranking a relatively small candidate set ($N=20$).
Our framework limits MLLM inference to a bounded candidate list rather than applying the model to the entire item catalog. Additionally, concatenating item images into composite visual inputs simplifies multimodal prompt construction by reducing the number of separately supplied images, although the actual visual-token cost depends on the model's image processing strategy. Nevertheless, the observed inference latency remains a substantial challenge for real-time recommendation. More efficient MLLMs, optimized inference, or offline reranking may help improve practicality, but evaluating these alternatives is beyond the scope of this study.

\section{Conclusion}
\label{sec:conclusion}
In this work, we present \ours, a training-free framework for diversity-aware recommendation reranking using pretrained MLLMs.
Experiments across product and video recommendation domains reveal three key findings: (1)~the benefits of visual information are domain-dependent, with substantial gains in video recommendation but limited improvements on product datasets, (2)~MLLM-based reranking can simultaneously improve recommendation relevance and category-diversity-sensitive ranking quality without task-specific training and (3)~the magnitude of visual gains and the consistency of reranking improvements depend on the MLLM backbone, with larger visual gains not necessarily yielding a stronger reranker.
Our results demonstrate the potential of pretrained MLLMs to incorporate multimodal information and diversity-aware instructions into existing recommendation pipelines, while highlighting differences across domains and model backbones.
Future work includes fine-tuning MLLMs for recommendation-specific objectives, exploring user-controllable diversity preferences, and improving inference efficiency for larger-scale reranking.

\section*{Ethical Considerations}
This work aims to reduce repetitive, narrow recommendations, which can mitigate filter-bubble effects; however, diversity measured by category coverage may not reflect what individual users value, and diversity objectives should not override user intent. Pretrained MLLMs may encode social biases from their training data, and relying on visual appearance (e.g., cover art) could amplify stereotypes or favor items with professionally produced imagery, disadvantaging smaller creators or sellers; we recommend auditing exposure across item groups before deployment. All experiments on the proprietary Prime Video data were conducted within internal company infrastructure: InternVL3-38B was run on local GPUs, and Claude Sonnet 4.6 was accessed through an internal enterprise API under which user data is not exposed externally and is not used for model training. We report only relative improvements on this dataset and release no user-level data.
Finally, large MLLM inference is energy-intensive; restricting reranking to small candidate sets limits, but does not remove, this cost.

\bibliographystyle{ACM-Reference-Format}
\bibliography{ref}

@inproceedings{liu2023visual,
  title={Visual Instruction Tuning},
  author={Liu, Haotian and Li, Chunyuan and Wu, Qingyang and Lee, Yong Jae},
  booktitle={Advances in Neural Information Processing Systems},
  volume={36},
  pages={34892--34916},
  year={2023}
}

@inproceedings{chen2022denoising,
  title={Denoising self-attentive sequential recommendation},
  author={Chen, Huiyuan and Lin, Yusan and Pan, Menghai and Wang, Lan and Yeh, Chin-Chia Michael and Li, Xiaoting and Zheng, Yan and Wang, Fei and Yang, Hao},
  booktitle={Proceedings of the 16th ACM conference on recommender systems},
  pages={92--101},
  year={2022}
}

@inproceedings{alayrac2022flamingo,
  title={Flamingo: A Visual Language Model for Few-Shot Learning},
  author={Alayrac, Jean-Baptiste and Donahue, Jeff and Luc, Pauline and Miech, Antoine and Barr, Iain and Hasson, Yana and Lenc, Karel and Mensch, Arthur and Millican, Katherine and Reynolds, Malcolm and others},
  booktitle={Advances in Neural Information Processing Systems},
  volume={35},
  pages={23716--23736},
  year={2022}
}

@inproceedings{zhao2024can,
  title={Can one embedding fit all? a multi-interest learning paradigm towards improving user interest diversity fairness},
  author={Zhao, Yuying and Xu, Minghua and Chen, Huiyuan and Chen, Yuzhong and Cai, Yiwei and Islam, Rashidul and Wang, Yu and Derr, Tyler},
  booktitle={Proceedings of the ACM web conference 2024},
  pages={1237--1248},
  year={2024}
}

@inproceedings{chen2021structured,
  title={Structured graph convolutional networks with stochastic masks for recommender systems},
  author={Chen, Huiyuan and Wang, Lan and Lin, Yusan and Yeh, Chin-Chia Michael and Wang, Fei and Yang, Hao},
  booktitle={Proceedings of the 44th international ACM SIGIR conference on research and development in information retrieval},
  year={2021}
}

@inproceedings{chen2024internvl,
  title={InternVL: Scaling Up Vision Foundation Models and Aligning for Generic Visual-Linguistic Tasks},
  author={Chen, Zhe and Wu, Jiannan and Wang, Wenhai and Su, Weijie and Chen, Guo and Xing, Sen and Zhong, Muyan and Zhang, Qinglong and Zhu, Xizhou and Lu, Lewei and others},
  booktitle={Proceedings of the IEEE/CVF Conference on Computer Vision and Pattern Recognition},
  pages={24185--24198},
  year={2024}
}

@inproceedings{liu2026rethinking,
  title={Rethinking Semantic ID Construction for Generative Recommendation: SimHash with Parallel Decoding and Semantic Alignment},
  author={Yuqing Liu and Huiyuan Chen and Yibo Wang and Wooseong Yang and Philip S. Yu},
  booktitle={Advances in Neural Information Processing Systems},
  year={2026}
}

@inproceedings{wei2019mmgcn,
  title={MMGCN: Multi-Modal Graph Convolution Network for Personalized Recommendation of Micro-Video},
  author={Wei, Yinwei and Wang, Xiang and Nie, Liqiang and He, Xiangnan and Hong, Richang and Chua, Tat-Seng},
  booktitle={Proceedings of the 27th ACM International Conference on Multimedia},
  pages={1437--1445},
  year={2019}
}

@inproceedings{he2016vbpr,
  title={VBPR: Visual Bayesian Personalized Ranking from Implicit Feedback},
  author={He, Ruining and McAuley, Julian},
  booktitle={Proceedings of the AAAI Conference on Artificial Intelligence},
  volume={30},
  number={1},
  year={2016}
}

@article{wang2020personalized,
  title={Personalized Re-ranking for Improving Diversity in Live Recommender Systems},
  author={Wang, Yichao and Zhang, Xiangyu and Liu, Zhirong and Dong, Zhenhua and Feng, Xinhua and Tang, Ruiming and He, Xiuqiang},
  journal={arXiv preprint arXiv:2004.06390},
  year={2020}
}

@article{wei2024towards,
  title={Towards Unified Multi-Modal Personalization: Large Vision-Language Models for Generative Recommendation and Beyond},
  author={Wei, Tianxin and Jin, Bowen and Li, Ruirui and Zeng, Hansi and Wang, Zhengyang and Sun, Jianhui and Yin, Qingyu and Lu, Hanqing and Wang, Suhang and He, Jingrui and others},
  journal={arXiv preprint arXiv:2403.10667},
  year={2024}
}

@inproceedings{kang2018self,
  title={Self-Attentive Sequential Recommendation},
  author={Kang, Wang-Cheng and McAuley, Julian},
  booktitle={2018 IEEE International Conference on Data Mining (ICDM)},
  pages={197--206},
  year={2018},
  organization={IEEE}
}

@inproceedings{li2023blip,
  title={BLIP-2: Bootstrapping Language-Image Pre-Training with Frozen Image Encoders and Large Language Models},
  author={Li, Junnan and Li, Dongxu and Savarese, Silvio and Hoi, Steven},
  booktitle={International Conference on Machine Learning},
  pages={19730--19742},
  year={2023},
  organization={PMLR}
}

@inproceedings{sun2023chatgpt,
  title={Is ChatGPT Good at Search? Investigating Large Language Models as Re-Ranking Agents},
  author={Sun, Weiwei and Yan, Lingyong and Ma, Xinyu and Wang, Shuaiqiang and Ren, Pengjie and Chen, Zhumin and Yin, Dawei and Ren, Zhaochun},
  booktitle={Proceedings of the 2023 Conference on Empirical Methods in Natural Language Processing},
  pages={14918--14937},
  year={2023}
}

@inproceedings{hou2024large,
  title={Large Language Models are Zero-Shot Rankers for Recommender Systems},
  author={Hou, Yupeng and Zhang, Junjie and Lin, Zihan and Lu, Hongyu and Xie, Ruobing and McAuley, Julian and Zhao, Wayne Xin},
  booktitle={European Conference on Information Retrieval},
  pages={364--381},
  year={2024},
  organization={Springer}
}

@misc{anthropic2024claude,
  title={The Claude Model Card},
  author={Anthropic},
  year={2024},
  howpublished={\url{https://www.anthropic.com}}
}

@article{carraro2024enhancing,
  title={Enhancing Recommendation Diversity by Re-ranking with Large Language Models},
  author={Carraro, Diego and Bridge, Derek},
  journal={ACM Transactions on Recommender Systems},
  year={2024},
  publisher={ACM}
}

@inproceedings{gao2025llm4rerank,
  title={LLM4Rerank: LLM-based Auto-Reranking Framework for Recommendations},
  author={Gao, Jingtong and Zheng, Bo and Qiu, Wenjie and Jia, Aobo and Zuo, Minghui and Tang, Ruiming and He, Xiangnan},
  booktitle={Proceedings of the ACM Web Conference},
  year={2025}
}

@inproceedings{he2025msrbench,
  title={When Large Vision Language Models Meet Multimodal Sequential Recommendation},
  author={He, Jiangyuan and others},
  booktitle={Proceedings of the ACM Web Conference},
  year={2025}
}

@inproceedings{ye2025mllmmsr,
  title={Harnessing Multimodal Large Language Models for Multimodal Sequential Recommendation},
  author={Ye, Yuyang and Zheng, Zhi and Shen, Yishan and Wang, Tianshu and Zhang, Hengruo and Zhu, Peijun and Yu, Runlong and Zhang, Kai and Xiong, Hui},
  booktitle={Proceedings of the AAAI Conference on Artificial Intelligence},
  volume={39},
  number={12},
  pages={13069--13077},
  year={2025}
}

@article{zhu2025internvl3,
  title={InternVL3: Exploring Advanced Training and Test-Time Recipes for Open-Source Multimodal Models},
  author={Zhu, Jinguo and Wang, Weiyun and Chen, Zhe and others},
  journal={arXiv preprint arXiv:2504.10479},
  year={2025}
}

@inproceedings{carbonell1998mmr,
  title={The Use of {MMR}, Diversity-Based Reranking for Reordering Documents and Producing Summaries},
  author={Carbonell, Jaime and Goldstein, Jade},
  booktitle={Proceedings of the 21st Annual International ACM SIGIR Conference on Research and Development in Information Retrieval},
  pages={335--336},
  year={1998}
}

@inproceedings{clarke2008novelty,
  title={Novelty and Diversity in Information Retrieval Evaluation},
  author={Clarke, Charles L. A. and Kolla, Maheedhar and Cormack, Gordon V. and Vechtomova, Olga and Ashkan, Azin and B{\"u}ttcher, Stefan and MacKinnon, Ian},
  booktitle={Proceedings of the 31st Annual International ACM SIGIR Conference on Research and Development in Information Retrieval},
  pages={659--666},
  year={2008}
}

@inproceedings{steck2018calibrated,
  title={Calibrated Recommendations},
  author={Steck, Harald},
  booktitle={Proceedings of the 12th ACM Conference on Recommender Systems},
  pages={154--162},
  year={2018}
}

@article{wu2024survey,
  title={A Survey on Large Language Models for Recommendation},
  author={Wu, Likang and Zheng, Zhi and Qiu, Zhaopeng and Wang, Hao and Gu, Hongchao and Shen, Tingjia and Qin, Chuan and Zhu, Chen and Zhu, Hengshu and Liu, Qi and Xiong, Hui and Chen, Enhong},
  journal={World Wide Web},
  volume={27},
  number={5},
  pages={60},
  year={2024},
  publisher={Springer}
}

@inproceedings{bao2023tallrec,
  title={{TALLRec}: An Effective and Efficient Tuning Framework to Align Large Language Model with Recommendation},
  author={Bao, Keqin and Zhang, Jizhi and Zhang, Yang and Wang, Wenjie and Feng, Fuli and He, Xiangnan},
  booktitle={Proceedings of the 17th ACM Conference on Recommender Systems},
  pages={1007--1012},
  year={2023}
}

@inproceedings{geng2022recommendation,
  title={Recommendation as Language Processing ({RLP}): A Unified Pretrain, Personalized Prompt \& Predict Paradigm ({P5})},
  author={Geng, Shijie and Liu, Shuchang and Fu, Zuohui and Ge, Yingqiang and Zhang, Yongfeng},
  booktitle={Proceedings of the 16th ACM Conference on Recommender Systems},
  pages={299--315},
  year={2022}
}

@inproceedings{sun2019bert4rec,
  title={{BERT4Rec}: Sequential Recommendation with Bidirectional Encoder Representations from Transformer},
  author={Sun, Fei and Liu, Jun and Wu, Jian and Pei, Changhua and Lin, Xiao and Ou, Wenwu and Jiang, Peng},
  booktitle={Proceedings of the 28th ACM International Conference on Information and Knowledge Management},
  pages={1441--1450},
  year={2019}
}

@inproceedings{hidasi2016session,
  title={Session-based Recommendations with Recurrent Neural Networks},
  author={Hidasi, Bal{\'a}zs and Karatzoglou, Alexandros and Baltrunas, Linas and Tikk, Domonkos},
  booktitle={Proceedings of the 4th International Conference on Learning Representations},
  year={2016}
}

@inproceedings{zhou2023bootstrap,
  title={Bootstrap Latent Representations for Multi-modal Recommendation},
  author={Zhou, Xin and Zhou, Hongyu and Liu, Yong and Zeng, Zhiwei and Miao, Chunyan and Wang, Pengwei and You, Yuan and Jiang, Feijun},
  booktitle={Proceedings of the ACM Web Conference 2023},
  pages={845--854},
  year={2023}
}

@article{liu2024multimodal,
  title={Multimodal Recommender Systems: A Survey},
  author={Liu, Qidong and Hu, Jiaxi and Xiao, Yutian and Zhao, Xiangyu and Gao, Jingtong and Wang, Wanyu and Li, Qing and Tang, Jiliang},
  journal={ACM Computing Surveys},
  volume={56},
  number={12},
  pages={1--37},
  year={2024},
  publisher={ACM}
}

@inproceedings{pei2019personalized,
  title={Personalized Re-ranking for Recommendation},
  author={Pei, Changhua and Zhang, Yi and Zhang, Yongfeng and Sun, Fei and Lin, Xiao and Sun, Hanxiao and Wu, Jian and Jiang, Peng and Ge, Junfeng and Ou, Wenwu and Pei, Dan},
  booktitle={Proceedings of the 13th ACM Conference on Recommender Systems},
  pages={3--11},
  year={2019}
}

@inproceedings{chen2018fast,
  title={Fast Greedy {MAP} Inference for Determinantal Point Process to Improve Recommendation Diversity},
  author={Chen, Laming and Zhang, Guoxin and Zhou, Eric},
  booktitle={Advances in Neural Information Processing Systems},
  volume={31},
  pages={5627--5638},
  year={2018}
}
\end{document}